\documentclass{PoS}

\usepackage{booktabs}
\usepackage{amsmath}

\title{Disconnected Loop Subtraction Methods in Lattice QCD}

\ShortTitle{Disconnected Loop Subtraction Methods in Lattice QCD}

\author{\speaker{Travis Whyte} \\
        Department of Physics, Baylor University, Waco, Texas, 76798-7316, USA\\
        E-mail: \email{travis\_whyte@baylor.edu}}

\author{Suman Baral\\
        Everest Institute of Science and Technology, Samakhusi Kathmandu, Nepal\\
        E-mail: \email{suman@everestsciencetech.com}}
        
\author{Paul Lashomb \\
        Department of Physics, Baylor University, Waco, Texas, 76798-7316, USA\\
        E-mail: \email{paul\_lashomb@baylor.edu}}

\author{Walter Wilcox \\
        Department of Physics, Baylor University, Waco, Texas, 76798-7316, USA\\
        E-mail: \email{walter\_wilcox@baylor.edu}}
        
\author{Ronald B. Morgan\\
        Department of Mathematics, Baylor University, Waco, TX 76798-7328, USA\\
        E-mail: \email{ronald\_morgan@baylor.edu}}

\abstract{Noise subtraction methods are a set of techniques that aim to reduce the variance of signals in LQCD which are often flooded with noise. The standard approach is a pertubative subtraction. In this work, we demonstrate the abilities of our new noise subtraction methods with methods which show considerable improvement over pertubative subtraction in the reduction of the variance for the set of LQCD operators that were studied. The methods were tested at $\kappa_{crit}$ on quenched configurations, as well as on dynamical quark configurations at $\kappa = 0.1453$. A significant improvement in the reduction of operator variance was observed in both cases.}

\FullConference{37th International Symposium on Lattice Field Theory - Lattice2019\\
		16-22 June 2019\\
		Wuhan, China}

\begin{document}

\section{Introduction}

Physical quantities, represented as operators in Lattice QCD, tend to produce signals with a large amount of noise. One way to combat this is through noise subtraction methods. We have previously \cite{baral} developed and implemented several new types of noise subtraction methods. Due to their large computational cost, approximations were made to avoid disconnected loop effects. In this work, these effects are included using a dynamical configuration in order to test the effectiveness of the methods discussed in \cite{baral19} when disconnected loop effects are included.

\section{Noise Subtraction Methods}
%In noise subtraction, given two matrices $Q$ and $\tilde{Q}$, where $\tilde{Q}$ is traceless, one can show that, 

%\begin{equation}
%    \langle Tr(QX) \rangle = \langle Tr\{(Q-\tilde{Q})\} 
%\end{equation}\nointent
%In other words, the expectation value of the trace remains unchanged by addition of a traceless matrix. 

%In LQCD, however, the expectation value of an operator can be written as, 

%\begin{equation}
%    \langle \bar{\psi} \Theta \bar{\psi} \rangle = -Tr(\Theta M^{-1})
%\end{equation}

The basis for noise subtraction methods involves using an important result of noise theory, namely, that, if one defines a matrix $X$ made from Z($N \geq$ 3) noise vectors $\eta^{(n)}$ as

\begin{equation}
    X_{ij} \equiv \frac{1}{N} \sum^N_n \eta^{(n)}_i \eta^{*(n)}_j   .
\end{equation}\noindent
Then, if $Q$ is the matrix representation of an operator, one can show that \cite{baral19}

\begin{equation} \label{eq:var}
    V[Tr(QX_{Z(N \geq 3)})] = \frac{1}{N} \sum_{i \neq j} |q_{ij}|^2.
\end{equation}
In other words, the variance of $Tr(QX_{Z(N \geq 3)})$ depends only on the off-diagonal elements of $Q$.

The foundation of noise subtraction techniques involves noticing that, for two matrices $Q$ and $\tilde{Q}$, such that $\tilde{Q}$ is traceless,

\begin{equation}
    \langle Tr(QX) \rangle = \Big \langle Tr \Big \{ (Q - \tilde{Q}) X \Big \} \Big \rangle
\end{equation}\noindent
and, from Eq. \ref{eq:var},

\begin{equation}  \label{eq:var2}
    V\Big [ Tr \Big \{ (Q - \tilde{Q}) X_{Z(N \geq 3)} \Big \} \Big ] = \frac{1}{N} \sum_{i \neq j} (|q_{ij} - \tilde{q}_{ij} |^2).
\end{equation}
In LQCD, trace terms appear as the expectation value of operators, 
\begin{equation}
    \langle \bar{\psi} \Theta \psi \rangle = -Tr(\Theta M^{-1}).
\end{equation}\noindent
Through projecting the quark matrix onto the noise vectors $\eta^{(n)}$ as
\begin{equation}
    Mx^{(n)} = \eta^{(n)},
\end{equation}\noindent
and combining it with the result of Eq. \ref{eq:var2}, it can also be shown \cite{baral19} that the for $Z(4)$ noise, the variance of such a trace depends only on the off-diagonal elements of $M^{-1}$. This suggests that if one constructs a suitable traceless inverse to the QCD matrix, $\tilde{M}^{-1}$, such that its off-diagonal elements are similar to $M^{-1}$, then the variance of the operator $\Theta$ can be reduced by adding the approximate inverse to the inverse QCD matrix. 

The standard method for noise subtraction is perturbative subtraction (PS). Our new noise subtraction methods are referred to as eigenvalue subtraction (ES)\cite{guerrero}, polynomial subtraction (POLY)\cite{liu}, and Hermitian-forced eigenvalue subtraction (HFES)\cite{guerrero_dis}, each of which produced modest decreases in the variance of several operators. The most promising of the methods we introduced, however, involves a combination of these methods. These combination methods come about by combining the techniques of HFES with POLY and PS, referred to as HFPS and HFPOLY, respectively. For the HFPOLY method, a general trace $Tr(\Theta M^{-1})$ involving some operator $\Theta$ and QCD matrix $M$ for random noise vectors $\eta^{(n)}$ of size $N$ in noise space can be shown \cite{baral19} to be given by
\begin{equation}
    \begin{aligned}
    Tr(\Theta M^{-1}) = & \frac{1}{N}\sum_n^N(
    \eta^{(n)\dagger}\Theta [x^{(n)}  - \tilde{x}'^{(n)}_{eig} - (\tilde{x}^{(n)}_{poly} - \tilde{x}'^{(n)}_{eigpoly}) ])
    + Tr(\Theta\gamma_5\tilde{M}'^{-1}_{eig})\\ 
    & + Tr(\Theta\tilde{M}'^{-1}_{poly} - \Theta\gamma_5\tilde{M}'^{-1}_{eigpoly})
    \end{aligned}
\end{equation}\noindent
where
\begin{gather}
        x^{(n)} = M^{-1} \eta^{(n)}, \\
        \tilde{x}^{(n)} \equiv \tilde{M}'^{-1}_{poly}\eta^{(n)},\\
        \tilde{x}'^{(n)}_{eig} \equiv \gamma_5\tilde{M}'^{-1}_{eig}\eta^{(n)}
        = \gamma_5\sum_q^Q \frac{1}{\lambda'(q)}e'^{(q)}_R(e'^{(q)\dagger}_R\eta^{(n)}),\\
        \tilde{x}'^{(n)}_{eigpoly} \equiv \gamma_5\tilde{M}'^{-1}_{eigpoly}\eta^{(n)}
        = \gamma_5\sum_q^Q \frac{1}{\xi'(q)}e'^{(q)}_R(e'^{(q)\dagger}_R\eta^{(n)}).
\end{gather}
%\begin{equation}
%    \tilde{x}^{(n)} \equiv %\tilde{M}'^{-1}_{poly}\eta^{(n)}
%\end{equation}
%\begin{equation}
%    \tilde{x}'^{(n)}_{eig} \equiv %\gamma_5\tilde{M}'^{-1}_{eig}\eta^{(n)}
%    = \gamma_5\sum_q^Q \frac{1}{\lambda'(q)}e'^{(q)}_R(e'^{(q)\dagger}_R\eta^{(n)})
%\end{equation}
%\begin{equation}
%    \tilde{x}'^{(n)}_{eigpoly} \equiv %\gamma_5\tilde{M}'^{-1}_{eigpoly}\eta^{(n)}
%    = \gamma_5\sum_q^Q %\frac{1}{\xi'(q)}e'^{(q)}_R(e'^{(q)\dagger}_R\eta^{(n)})
%\end{equation}\noindent
$e'^{(q)}_R$ are the eigenmodes of $M'$ ($M' \equiv M\gamma_5$), and $\lambda'(q)$ and $\xi'(q)$ are the eigenvalues of $\tilde{M}'_{eig}$ and $\tilde{M}'_{eigpoly}$, respectively. 

The approximate QCD matrix inverses, $\tilde{M}'_{eig}$, $\tilde{M}_{poly}$, $\tilde{M}'_{eigpoly}$, and $\tilde{M}^{-1}_{pert}$ are given by
\begin{gather}
    \tilde{M}^{-1}_{pert} \equiv I + \kappa P + (\kappa P)^2 + (\kappa P)^3 + (\kappa P)^4 + (\kappa P)^5 + (\kappa P)^6, \\
    \tilde{M}^{-1}_{poly} \equiv b_0 I + b_1 \kappa P + b_2(\kappa P)^2 + b_3(\kappa P)^3 + b_4(\kappa P)^4 + b_5(\kappa P)^5 + b_6(\kappa P)^6, \\
    \tilde{M}'^{-1}_{eig} \equiv \tilde{V}'_R \tilde{\Lambda}'^{-1} \tilde{V}'^{\dagger}_R,\\
    \tilde{M}'^{-1}_{eigpoly} \equiv \tilde{V}'_R \tilde{\Xi}'^{-1} \tilde{V}'^{\dagger}_R,
\end{gather}\noindent
%\begin{equation}
%    \tilde{M}'^{-1}_{eigpoly} \equiv \tilde{V}'_R %\tilde{\Xi}'^{-1} \tilde{V}'^{\dagger}_R 
%\end{equation}\noindent
where $P$ is the quark hopping matrix, $\kappa$ is the hopping parameter, $\tilde{V}'$ is a matrix whose columns are the $Q$ smallest right eigenvalues of $M'$, and $\tilde{\Lambda}'^{-1}$ and $\tilde{\Xi}'^{-1}$ 
are diagonal matrices that contain along their diagonal the reciprocal eigenvalues $1/\lambda'^{(q)}$ and $1/\xi'^{(q)}$, respectively. For the POLY and HFPOLY methods, the coefficients are found using the power basis method. In the case of the HFPS method, the approximate matrix inverse $\tilde{M}^{-1}_{poly}$ is replaced with $\tilde{M}^{-1}_{pert}$. A comprehensive description of each of the methods can be found in Ref. \cite{baral19}. 

\section{Results}

\subsection{Quenched Configurations at $\kappa_{crit}$}

\begin{figure}[h]
    \centering
    \includegraphics[trim={5cm 8.5cm 5cm 8.5cm },width=.45\textwidth]{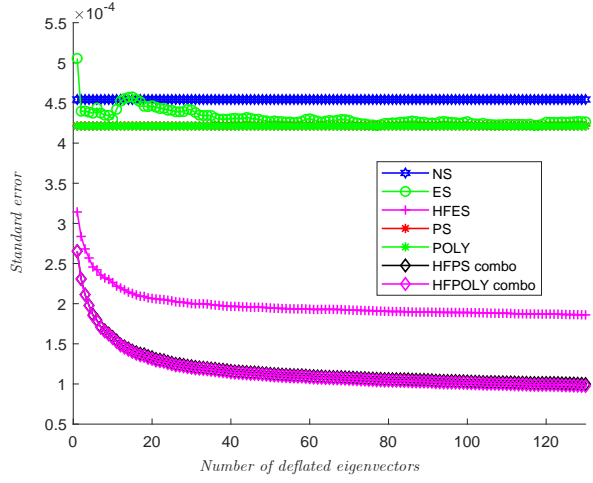}
    \caption{Error bars of operator as a function of deflated eigenvalues on the quenched configurations at $\kappa_{crit}$ for the local vector: $Im[\bar{\psi}(x)\gamma_\mu\psi(x)]$.}
    \label{fig:fig_1}
\end{figure}

\begin{figure}[h]
    \centering
    \includegraphics[trim={5cm 8.5cm 5cm 8.5cm },width=.45\textwidth]{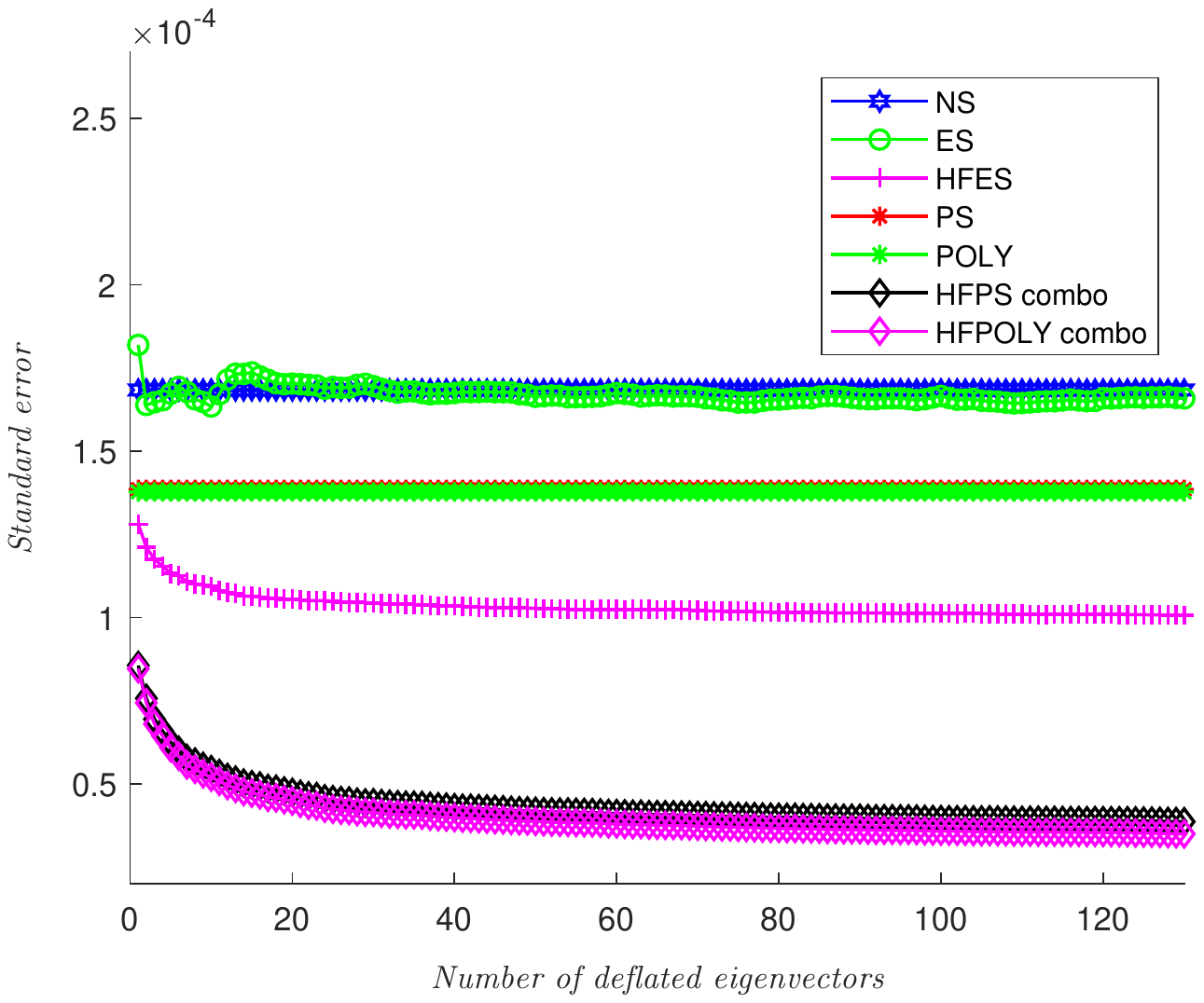}
    \caption{Error bars of operator as a function of deflated eigenvalues on the quenched configurations at $\kappa_{crit}$ for the point-split vector: $\kappa Im[\bar{\psi}(x+a_\mu)(1+\gamma_{\mu})U^{\dagger}_{\mu} \psi(x)] - \kappa Im[\bar{\psi}(x)(1-\gamma_{\mu})U_\mu \psi(x+a_{\mu})]$.}
    \label{fig:fig_2}
\end{figure}

\begin{figure}[h]
    \centering
    \includegraphics[trim={5cm 8.5cm 5cm 8.5cm },width=.45\textwidth]{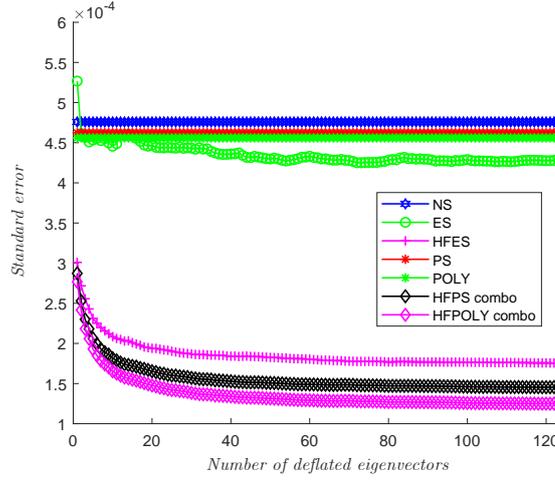}
    \caption{Error bars of operator as a function of deflated eigenvalues on the quenched configurations at $\kappa_{crit}$ for the scalar: $Re[\bar{\psi}(x)\psi(x)]$.}
    \label{fig:fig_3}
\end{figure}

The dimensions of the quenched lattice were $24^3 \times 32$ and were run at $\beta = 6.0$ with a hopping parameter of $\kappa = 0.1570 \approx \kappa_{crit}$. The standard error was averaged over 10 configurations. Plots of the error bars versus the number of deflated eigenvalues are shown in Figure \ref{fig:fig_1}, Figure \ref{fig:fig_2}, and Figure \ref{fig:fig_3} corresponding to a local vector, point-split vector, and scalar operators, respectively. The methods compared are non-subtraction (NS), ES, PS, POLY, HFES, HFPS, and HFPOLY. In each case, the methods HFPS and HFPOLY produce a significant reduction in the error bars over both NS and PS.

% Kappa critical Local operator

% Kappa critical Point Split operator

% Kappa critical Scalar operator

Comparisons between the methods were made by defining the relative efficiencies, $RE$, as,

\begin{equation}
 RE \equiv (\frac{1}{\delta y^2} - 1 ) \times 100 \%   
\end{equation}
where $\delta y^2$ is the relative variance. The $RE$ of POLY, HFES, HFPS, and HFPOLY both to NS and to PS are shown in Table \ref{table:table_1}. For every operator, HFPOLY and HFPS produced significant improvements over PS in reduction of the error bars. 
% \usepackage{booktabs}
\begin{table}
    \centering
    \begin{tabular}{@{}lllllll@{}}
        \toprule
         & \multicolumn{2}{c}{\textbf{Scalar}} & \multicolumn{2}{c}{\textbf{Local}} & \multicolumn{2}{c}{\textbf{Point-Split}} \\ \midrule\midrule
         & \textbf{vs. NS} & \textbf{vs. PS} & \textbf{vs. NS} & \textbf{vs. PS} & \textbf{vs. NS} & \textbf{vs. PS} \\ \midrule
        \textbf{POLY} & 8.9\% & 2.8\% & 16.4\% & 0.1\% & 49.5\% & 1.1\% \\
        \textbf{HFES} & 634\% & 593\% & 496\% & 413\% & 180\% & 89.2\% \\
        \textbf{HFPS} & 972\% & 911\% & 1970\% & 1680\% & 1800\% & 1180\% \\
        \textbf{HFPOLY} & 1350\% & 1270\% & 2070\% & 1770\% & 2200\% & 1470\% \\ \bottomrule
    \end{tabular}
    \caption{Relative efficiencies for the quenched configurations at $\kappa_{crit}$}
    \label{table:table_1}
\end{table}

\subsection{Dynamical Configurations}

\begin{figure}[h]
    \centering
    \includegraphics[trim={5cm 8.5cm 5cm 8.5cm },width=.45\textwidth]{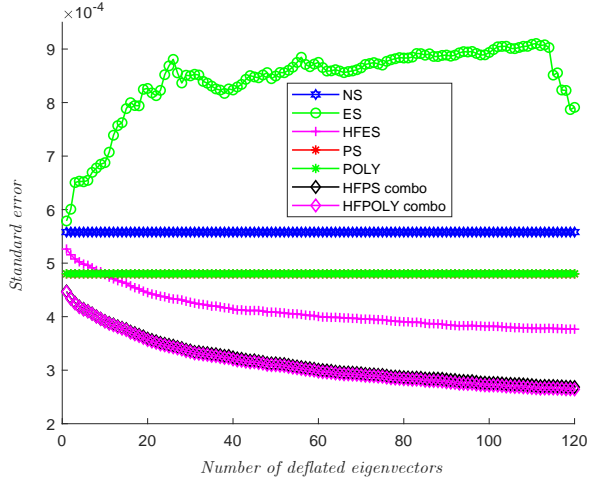}
    \caption{Error bars of operator as a function of deflated eigenvalues on the dynamical configurations at $\kappa = 0.1453$ for the local vector: $Im[\bar{\psi}(x)\gamma_\mu\psi(x)]$.}
    \label{fig:fig_4}
\end{figure}

\begin{figure}[h]
    \centering
    \includegraphics[trim={5cm 8.5cm 5cm 8.5cm },width=.45\textwidth]{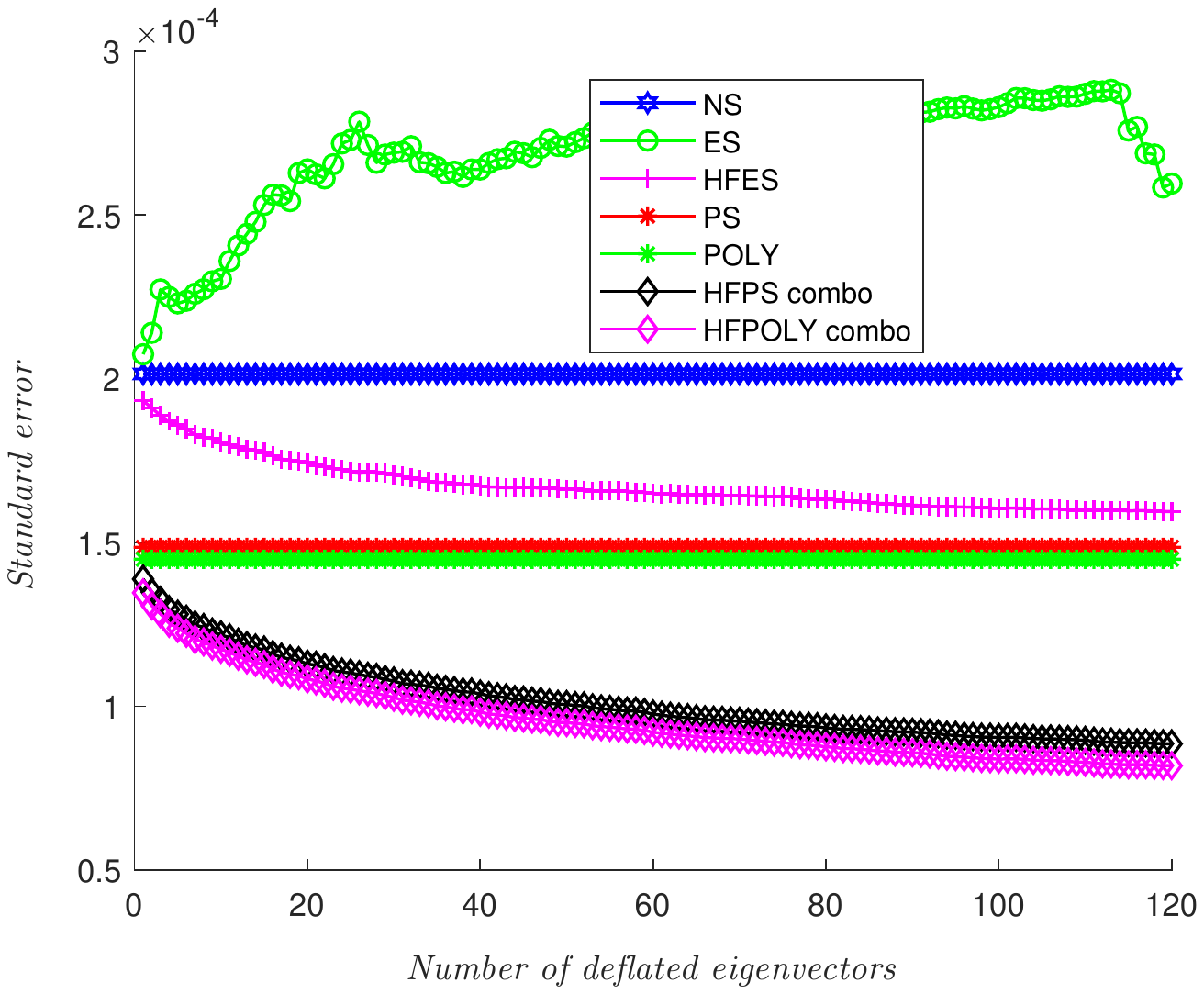}
    \caption{Error bars of operator as a function of deflated eigenvalues on the dynamical configurations at $\kappa = 0.1453$ for the point-split vector: $\kappa Im[\bar{\psi}(x+a_\mu)(1+\gamma_{\mu})U^{\dagger}_{\mu} \psi(x)] - \kappa Im[\bar{\psi}(x)(1-\gamma_{\mu})U_\mu \psi(x+a_{\mu})]$.}
    \label{fig:fig_5}
\end{figure}

\begin{figure}[h]
    \centering
    \includegraphics[trim={5cm 8.5cm 5cm 8.5cm },width=.45\textwidth]{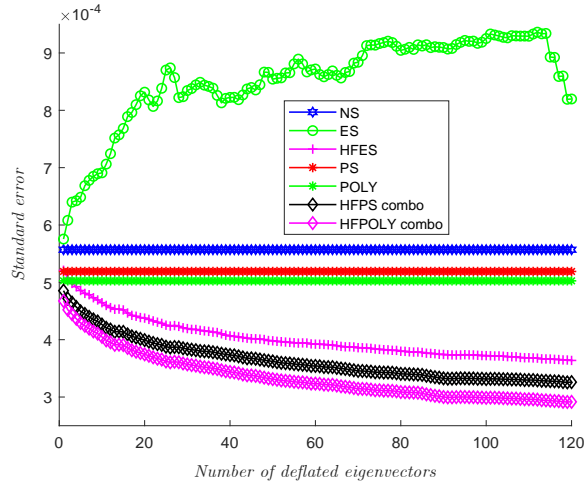}
    \caption{Error bars of operator as a function of deflated eigenvalues on the dynamical configurations at $\kappa = 0.1453$ for the scalar: $Re[\bar{\psi}(x)\psi(x)]$.}
    \label{fig:fig_6}
\end{figure}

The dynamical $N_f = 2 + 1 + 1$ MILC configurations consisted of a $16^3 \times 48$ lattice with $\beta = 5.8$ and a pion mass of $m_\pi = 306.9(5)$ MeV. Using 10 configurations, analysis of the pion correlator produced a hopping parameter of $\kappa = 0.1453$. This corresponds to the quenched case with an approximate hopping parameter $\kappa \approx 0.1567$ \cite{baral19}. Shown in Figure \ref{fig:fig_4}, Figure \ref{fig:fig_5}, and Figure \ref{fig:fig_6} are comparisons of the error bars of each of the three operators as a function of the deflated eigenvalues for the different methods. Notice, once again, the significant improvement in $RE$ for HFPS and HFPOLY over PS alone, also shown in Table \ref{table:table_2}. Due to deflation, an even better reduction in the variance is expected as the pion mass decreases closer to the physical point.

% Dynamical Local operator

% Dynamical Point Split

% Dynamical Scalar

% \usepackage{booktabs}
\begin{table}
    \centering
    \begin{tabular}{@{}lllllll@{}}
        \toprule
         & \multicolumn{2}{c}{\textbf{Scalar}} & \multicolumn{2}{c}{\textbf{Local}} & \multicolumn{2}{c}{\textbf{Point-Split}} \\ \midrule\midrule
         & \textbf{vs. NS} & \textbf{vs. PS} & \textbf{vs. NS} & \textbf{vs. PS} & \textbf{vs. NS} & \textbf{vs. PS} \\ \midrule
        \textbf{POLY} & 22.8\% & 6.6\% & 35.0\% & -0.1\% & 93.4\% & 5.2\% \\
        \textbf{HFES} & 134\% & 104\% & 120\% & 62.4\% & 60.0\% & -13.2\% \\
        \textbf{HFPS} & 192\% & 153\% & 332\% & 220\% & 417\% & 181\% \\
        \textbf{HFPOLY} & 260\% & 217\% & 436\% & 230\% & 505\% & 229\% \\ \bottomrule
    \end{tabular}
    \caption{Relative efficiencies for the dynamical configurations}
    \label{table:table_2}
\end{table}

\section{Conclusions}
Using the eigenmodes of the Hermitian Wilson matrix in the HFPS and HFPOLY methods resulted in a large variance reduction compared to the PS method alone near zero quark mass,
an improvement which is seen in both the quenched configurations at $\kappa_{crit}$ as well as the dynamical configurations at $\kappa = 0.1453$. As pion mass decreases towards 
the physical point, we expect even better reduction in the variance from deflation. There are still other areas of exploration including examining different LQCD operators, 
applying POLY to solving the linear systems rather than simply in deflation, and considering a POLY expansion for the Hermitian system. 

\section{Acknowledgements}
Numerical work was performed using the High Performance Cluster at Baylor University. We would like to thank the Baylor University Research Committee and the Texas Advanced Super Computing Center for partial support. We also would like to thank Abdou Abdel-Rehim and Victor Guerrero for their contributions to this work. We also would like to thank Doug Toussaint, Carlton DeTar, and Jim Hetrick for their help in obtaining and analyzing the MILC configurations. We also thank Randy Lewis for his QQCD program.

\end{document}